\documentclass[journal,transmag]{IEEEtran}
\usepackage{hyperref,color,multirow}
\usepackage{geometry}
\usepackage{graphicx,color,epsf}
\usepackage{amsmath,amsfonts,amssymb,amscd,bm}
\usepackage{siunitx}
\usepackage{pgfplots}
\usepackage{subfigure}
\usepackage{siunitx}
\usepackage[english]{babel}
\usepackage[utf8]{inputenc}
\usepackage[ruled,norelsize,linesnumbered]{algorithm2e}
\makeatletter
\newcommand{\removelatexerror}{\let\@latex@error\@gobble}
\makeatother
\newcommand{\numcpu}{N}
\newcommand{\im}{\mathbf{i}}

\usepackage{eso-pic}

\AddToShipoutPicture*{\footnotesize\sffamily\raisebox{0.6in}{\hspace{0.65in}\fbox{\parbox{\textwidth}{\copyright~2017
IEEE. Personal use of this material is permitted. Permission from IEEE
must be obtained for all other uses, in any current or future media,
including reprinting/republishing this material for advertising or
promotional purposes, creating new collective works, for resale or
redistribution to servers or lists, or reuse of any copyrighted
component of this work in other works.}}}}

\begin{document}
\newgeometry{top=0.7in, left=0.65in, textheight=9.2in, textwidth=7.2in}

\title{Parallel-In-Time Simulation of Eddy Current Problems Using Parareal}

\author{\IEEEauthorblockN{
	Sebastian Schöps\IEEEauthorrefmark{1},
	Innocent Niyonzima\IEEEauthorrefmark{2}, and
	Markus Clemens\IEEEauthorrefmark{3}}

	\IEEEauthorblockA{
		\IEEEauthorrefmark{1}Technische Universit\"{a}t Darmstadt, Institut f\"{u}r Theorie Elektromagnetischer Felder,\\Schlossgartenstrasse 8, D-64289 Darmstadt, Germany
		}

	\IEEEauthorblockA{
		\IEEEauthorrefmark{2}Columbia University, Civil Engineering and Engineering Mechanics,\\ 500 West 120th Street, New York, NY 10027, USA
	}

	\IEEEauthorblockA{
	\IEEEauthorrefmark{3}University of Wuppertal, Chair of Electromagnetic Theory,\\ Rainer-Gr\"{u}nter-Strasse 21, D-42119 Wuppertal, Germany
	}%
}%

\IEEEtitleabstractindextext{%
\begin{abstract}
In this contribution the usage of the Parareal method is proposed for the time-parallel solution of the eddy current problem. The method is adapted to the particular challenges of the problem that are related to the differential algebraic character due to non-conducting regions. It is shown how the necessary modification can be automatically incorporated by using a suitable time stepping method. The paper closes with a first demonstration of a simulation of a realistic four-pole induction machine model using Parareal. 
\end{abstract}

\begin{IEEEkeywords}
Parallel-in-time, eddy currents, time stepping, finite element
\end{IEEEkeywords}}

\maketitle
\thispagestyle{empty}
\pagestyle{empty}

\section{Introduction}
\IEEEPARstart{T}{he} numerical simulation of the eddy current problem in time domain is computationally 
expensive due to implicit time stepping. This is particularly challenging if long time periods have to be considered as for example when the start-up of an electrical machine is simulated, possibly with surrounding circuitry \cite{Tsukerman_2002aa}. Most implementations, as in GetDP \cite{Geuzaine_2007aa}, use low-order time stepping schemes as for example the $\theta$-method, \cite{Hairer_1996aa}. On the other hand, higher order time integration methods, e.g. \cite{Nicolet_1996aa,Benderskaya_2005aa}, have been proposed but are rarely used in practice. Recently, explicit methods gained interest as computational hardware architectures seem to favor those algorithms \cite{Auserhofer_2009ab, Schops_2012aa, Dutine_2017ab}. Another approach is time domain parallelization \cite{Gander_2015ab}. For example the reformulation of the time stepping process as one big system of equations has been proposed in \cite{Takahashi_2013aa}. However, it requires to rewrite the time-stepping code.

In this contribution the usage of the (non-intrusive) Parareal method \cite{Lions_2001aa,Gander_2008aa} is proposed and its application to a real world electrical engineering problem, i.e., an electrical machine, is shown. Furthermore, the method is adapted to the particular challenges of space discretized eddy current problems related to the differential algebraic character of the equation. The Parareal method has already been applied to wide range of problems in mathematics and in physics. These problems include for example linear and nonlinear parabolic problems 
\cite{Liu_2012aa}, molecular dynamics \cite{Baffico_2002aa}, 
stochastic differential equations \cite{Engblom_2009aa}, Navier Stokes \cite{Trindade_2004aa}, quantum control \cite{Maday_2007aa}. 

This paper is structured as follows: after this introduction, Section~\ref{sec:model} discusses the modeling and the differential algebraic character of the system. Section~\ref{sec:parareal} introduces the Parareal algorithm and its adaption. Finally, the numerical results are presented in Section~\ref{sec:num}. Conclusions are given in Section~\ref{sec:conclusion}.

\section{Modeling and Discretization}\label{sec:model}
\begin{figure}
	\centering
	\includegraphics[width=.24\textwidth]{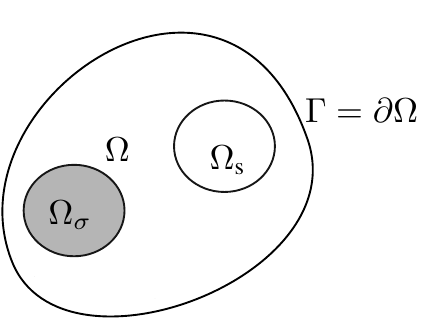}
	\vspace{-0.5em}
  \caption{Sketch of the computational domain of an eddy current problem}
  \label{fig:pics_patata}
\end{figure}

When disregarding displacement currents, and introducing the magnetic vector potential $\vec{A}$ as unknown, one obtains the eddy current problem in A$^\star$-formulation 
\begin{align}
	\sigma\partial_t{\vec{A}}
	+
	\nabla \times (\nu\nabla\times\vec{A})
	=
	\vec{J}_\mathrm{s}(t)
	\label{eq:mqs1}
\end{align}
on the domain $\Omega\times\mathcal{I}$ and $\mathcal{I}:=(t_0,t_\text{end}]$, see Fig.~\ref{fig:pics_patata}. The problem is well posed when supplying a gauge condition, suitable boundary conditions, e.g. Dirichlet
\begin{align*}
	\vec{n}\times\vec{A}|_\Gamma=0 \text{ where } \Gamma=\partial\Omega
\end{align*}
and an initial value $\vec{A}(\vec{r},t_0)=\vec{A}_0(\vec{r})$ with $\vec{r}\in\Omega$. The material is described by the conductivity $\sigma$ and nonlinear reluctivity $\nu$; the current density $\vec{J}_{\mathrm{s}} = \sum_k\vec{\chi}_{\mathrm{s},k}\,i_k$ is given by stranded-conductor winding functions $\vec{\chi}_{\mathrm{s},k}$, which homogeneously distribute the currents $i_k$. 

\subsection{Discretization by Finite Elements}
\begin{figure}[t]
	\centering
	\includegraphics[width=.245\textwidth]{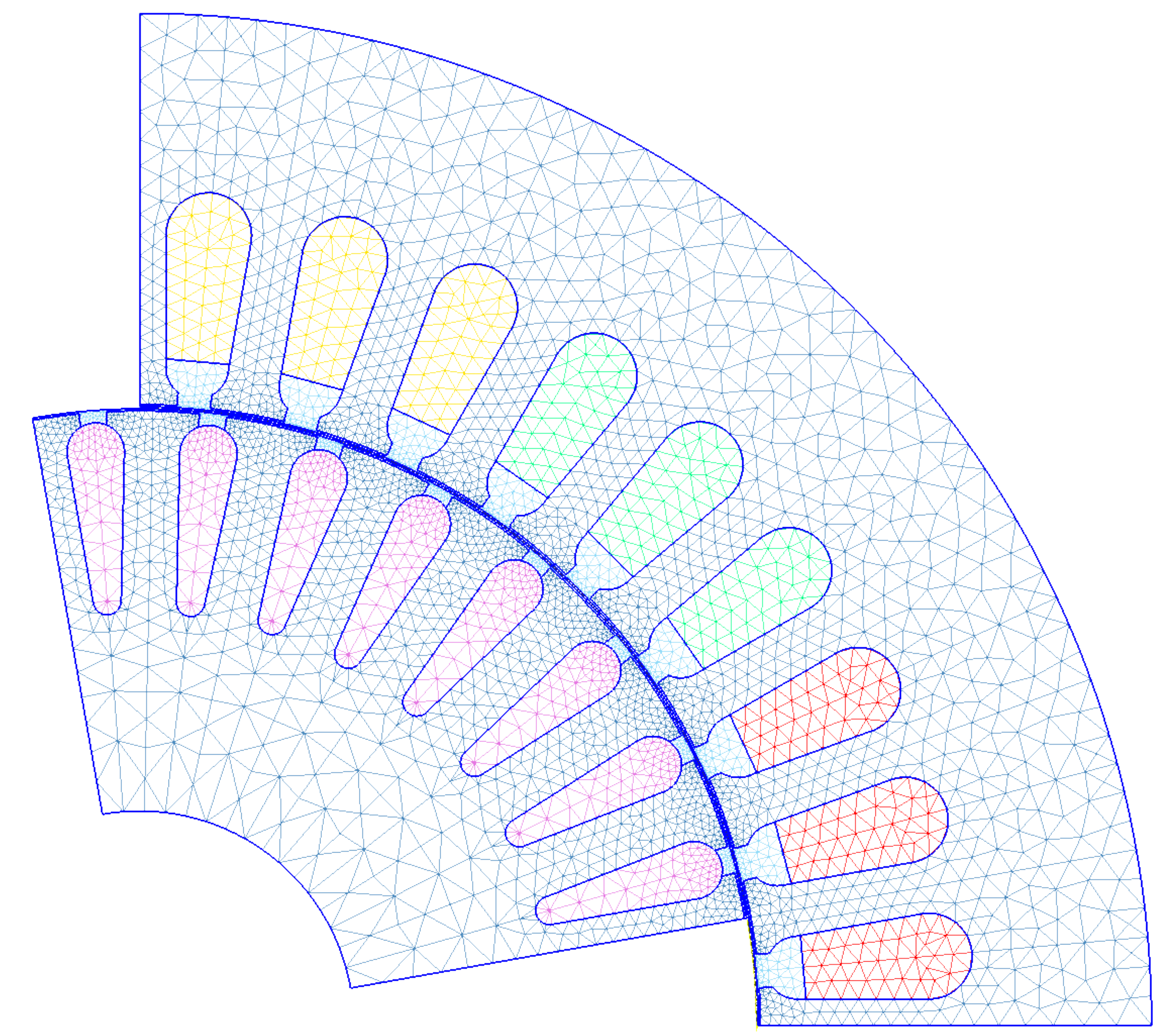}
	\caption{Mesh view of the four-pole induction machine model "im\_3kw" from the GetDP library \cite{Geuzaine_2007aa} as described in \cite{Gyselinck_2001aa}.}
	\label{fig:im_3kW}
\end{figure}

Equation \eqref{eq:mqs1} is reformulated in the following weak form: find $\vec{A}\in H_0(\mathrm{curl},\Omega)$ such that
\begin{align*}
	\int_\Omega \vec{w}\cdot\sigma\partial_t{\vec{A}}
	+
	\nabla\times\vec{w}\cdot(\nu\nabla\times\vec{A})\;\mathrm{d}\Omega
	&=
	\int_\Omega \vec{w}\cdot \vec{J}_{\mathrm{s}}\;\mathrm{d}\Omega
\end{align*}
\newgeometry{top=0.7in, left=0.65in, textheight=9.6in, textwidth=7.2in}
\noindent for all $\vec{w}\in H_0(\mathrm{curl},\Omega)$. Discretization by a finite set of edge elements \cite{Monk_2003aa} 
\begin{align*}
	\vec{A}(\vec{x},t)\approx\sum_{i=1}^{n}\vec{w}_i(\vec{x})\;a_i(t)
\end{align*}
yields for the induction machine "im\_3kw", cf. Fig.~\ref{fig:im_3kW}, the following system of differential algebraic equations (DAEs)
\begin{align}\label{eq:daes1}
	\mathbf{M}_\sigma\mathrm{d}_t \mathbf{a}(t)
	+
	\mathbf{K}_\mathbf{\nu}\bigl(\mathbf{a}(t),\theta(t)\bigr)\mathbf{a}(t)
	&=
	\mathbf{j}_{\mathrm{s}}(t)
\end{align}
where $\mathbf{a}(t)\in\mathbb{R}^{n}$ is the vector of (line-integrated) magnetic vector potentials, $\mathbf{M}_\sigma\in\mathbb{R}^{n\times n}$ denotes the (singular) mass matrix representing the conductivities and $\mathbf{j}_{\mathrm{s}}(t)\in\mathbb{R}^{n}$ describes the discretized source current density. Finally,
$\mathbf{K}_\nu(\mathbf{a},\theta)\in\mathbb{R}^{n\times n}$ is the curl-curl matrix which depends on rotor angle and flux. Movements are considered by the moving band approach \cite{Ferreira-da-Luz_2002aa} determined by the mechanical equation 
\begin{align}
  \label{eq:motion}
 \mathrm{d}_t\theta(t) = \omega(t)
 \quad\text{and}\quad
  I\mathrm{d}_t\omega(t)
  + \kappa\theta(t)
  =
 T\bigl(\mathbf{a}(t)\bigr)
\end{align}
with initial values $\theta(t_0)=\theta_0$ and $\omega(t_0)=\omega_0$, where $\omega$ is the angular velocity, $I$ the inertia, $\kappa$ the torsion coefficient and $T$ defines the mechanical excitation given by the magnetic field. 

Let us address in the following the current driven coupled problem \eqref{eq:daes1} and \eqref{eq:motion} as 
\begin{align}\label{eq:daes3}
	\mathbf{M}\mathrm{d}_t\mathbf{u}(t)
	+
	\mathbf{K}\bigl(\mathbf{u}(t))\mathbf{u}(t)
	&=
	\mathbf{f}(t)
\end{align}
with unknown $\mathbf{u}^{\!\top}=[\mathbf{a}^{\!\top},\theta,\omega]$ and the obvious definitions for $\mathbf{M}$, $\mathbf{K}$ and $\mathbf{f}$.

\subsection{Differential algebraic equations}
The solution of the DAE \eqref{eq:mqs1} is straightforward since the system is an index-1 DAE~\cite{Bartel_2011aa,Nicolet_1996aa}. It can be treated with standard techniques, while higher index problems are increasingly more difficult to solve \cite{Hairer_1996aa}.

Let $\mathbf{M}^+$ be the Moore-Penrose pseudo inverse of $\mathbf{M}$ such that $\mathbf{P}=\mathbf{M}^+\mathbf{M}$ and $\mathbf{Q}=\mathbf{I}-\mathbf{P}$ denote projectors decomposing the vector potential $\mathbf{u}_i=\mathbf{u}(t_i)$ at each time instance into its differential and algebraic components, respectively
\begin{align*}
	\mathbf{u}_i=\mathbf{P}\mathbf{u}_{i,\sigma}+\mathbf{Q}\mathbf{u}_{i,0}.
\end{align*}
When solving \eqref{eq:daes1} for given currents $\im_i=\im(t_i)$, 
only initial
conditions for the differential components $\mathbf{u}_{0,\sigma}$ may be prescribed. The algebraic components $\mathbf{u}_{0,0}$ must be consistently determined by solving the constraint
\begin{align}\label{eq:constraint}
	\mathbf{Q}^{\!\top}\mathbf{K}_\nu\mathbf{Q}\mathbf{u}_{0,0}
	=
	\mathbf{Q}^{\!\top}\mathbf{X}_{\mathrm{s}}\im_0
	-
	\mathbf{Q}^{\!\top}\mathbf{K}_\nu\mathbf{P}\mathbf{u}_{0,\sigma}.
\end{align}
However, when using the implicit Euler method to solve an initial value problem with inconsistent data, i.e, $\mathbf{u}_{0,\sigma}$ and $\mathbf{u}_{0,0}$ do not fulfill \eqref{eq:constraint}, a projection is automatically carried out: the time stepping instruction for $t_i$ to $t_{i+1}=t_i+\delta t$
\begin{align}\label{eq:euler}
	\Bigl(\frac{1}{\delta t}\mathbf{M}+\mathbf{K}\bigl(\mathbf{u}_{i+1}\bigr)\Bigr)\mathbf{u}_{i+1}
	=
	\mathbf{f}_{i+1}+\frac{1}{\delta t}\mathbf{M}\mathbf{u}_{i}
\end{align}
ignores inconsistent algebraic components after the first step due to the term $\mathbf{M}\mathbf{u}_{i}=\mathbf{M}\mathbf{P}\mathbf{u}_{i}=\mathbf{M}\mathbf{u}_{i,\sigma}$. This is generally not the case for higher index DAEs and other time-stepping schemes, see e.g. \cite{Bartel_2011aa}.

The implicit Euler method is a numerical implementation of the solution operator $\mathcal{F}:\mathcal{I}\times\mathcal{I}\times\mathbb{R}^{n} \rightarrow \mathbb{R}^{n}$ such that
\begin{align*}
	\mathbf{u}_i = \mathcal{F}(t_i, t_0, \mathbf{u}_0)
\end{align*}
which propagates $\mathbf{u}_0$ through time. Let us define another coarse propagator denoted by $\mathcal{G}:\mathcal{I}\times\mathcal{I}\times\mathbb{R}^{n} \rightarrow \mathbb{R}^{n}$ of lower precision, e.g., implicit Euler with a time step $\Delta t\gg\delta t$.
\begin{figure}
	\centering
	\includegraphics[width=.39\textwidth]{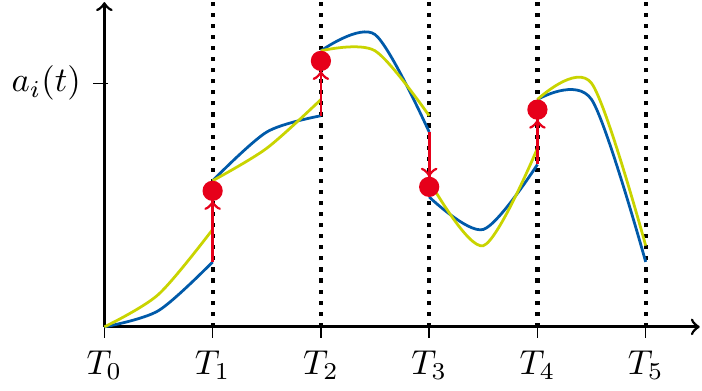}
	\caption{Jumps at the interfaces are compensated by Newton's method; coarse solution $\mathcal{G}$ (blue), correction via gradient (red) and fine solution $\mathcal{F}$ (green)}
	\label{fig:matching}
\end{figure}%
\section{The Parareal Method}\label{sec:parareal}
Let us split the total time interval into smaller intervals 
$\mathcal{I}_j:=(T_{j-1},T_{j}]$ with $t_0=T_0<T_1<\ldots<T_{\numcpu}=t_{\text{end}}$
according to the number of CPUs $\numcpu$ available. On each interval, one defines the equation
\eqref{eq:daes3} with initial value $\mathbf{U}_{j-1}:=\mathbf{u}(T_{j-1})$, 
final value $\mathbf{U}_{j}:=\mathbf{u}(T_{j})$. Continuity at the interfaces $T_j$ is established by matching conditions, see Fig.~\ref{fig:matching}
\begin{equation}
	\label{eq:newton}
	\mathbf{H}(\mathbf{U}):=\begin{cases}
		{\mathbf{U}_0} - \mathbf{u}_0 &= 0,
		\\
		{\mathbf{U}_1} - \mathcal{F}(T_1,T_0,{\mathbf{U}_{0}}) &= 0,
		\\
		&\vdots
		\\
		{\mathbf{U}_{\numcpu-1}} - \mathcal{F}(T_{\numcpu-2},T_{\numcpu-1}, {\mathbf{U}_{\numcpu-2}}) &= 0.
	\end{cases}
\end{equation}
In other words, the problem of matching can be considered as the unknown of a nonlinear equation $\mathbf{H}:\mathbb{R}^{N\cdot n}\to\mathbb{R}^{N\cdot n}$
in the variable ${\mathbf{U}}^{\!\top} = [\mathbf{U}_0^{\!\top},...,\mathbf{U}_j^{\!\top},...,\mathbf{U}_{\numcpu-1}^{\!\top}]$.
\subsection{Interpretation as Newton's Method}
The system \eqref{eq:newton} can be solved by Newton's method. It reads using the superscript $(k)$ to account for the iterations
\begin{equation}
	\label{eq:newton2}
	\frac{\partial \mathbf{H}}{\partial\mathbf{U}}\bigl(\mathbf{U}^{(k-1)}\bigr)\;
	(\mathbf{U}^{(k)} - \mathbf{U}^{(k-1)}) 
	= 
	-\mathbf{H}(\mathbf{U}^{(k-1)}).
\end{equation}
The $j$-th row of the Jacobian matrix $\partial \mathbf{H}/\partial \mathbf{U}$ has only two entries in columns $j-1$ and $j$. It is given by
\begin{equation}
	\frac{\partial H_j}{\partial\mathbf{U}}\bigl(\cdot\bigr)
	=
	\Bigl[\mathbf{0}, \ldots \mathbf{0}, 
	-\frac{\partial \mathcal{F}}{\partial\mathbf{U}}\bigl(T_j,T_{j-1},\cdot\bigr), 
	\mathbf{I} , \mathbf{0} \ldots,\mathbf{0} \Bigr]
\end{equation}
where $\mathbf{I}$ denotes the identity of dimension $n$.
After rearranging the terms, equation \eqref{eq:newton2} is equivalent to the explicit update formula
\begin{align}
	{\mathbf{U}_j^{(k)}}
	&= 
	\mathcal{F}(T_j, T_{j-1},{\mathbf{U}^{(k-1)}_{j-1}})\\
	&\qquad\quad+\frac{\partial\mathcal{F}}{\partial \mathbf{U}}\bigl(T_j, T_{j-1},\mathbf{U}_{j-1}^{(k-1)}\bigr)\; ( {\mathbf{U}^{(k)}_{j-1}} - {\mathbf{U}^{(k-1)}_{j-1}} ),\nonumber\\
	\intertext{in which the linearization is approximated by the difference}
	\label{eq:para}
	&\approx\mathcal{F}(T_j, T_{j-1},{\mathbf{U}^{(k-1)}_{j-1}})\\
	&\qquad\quad+
	\mathcal{G}(T_j, T_{j-1},{\mathbf{U}^{(k)}_{j-1}})
	-
	\mathcal{G}(T_j, T_{j-1},{\mathbf{U}^{(k-1)}_{j-1}})\nonumber
\end{align}
which is a Quasi-Newton method as proposed e.g. in \cite{Gander_2008aa}. 

Due to the splitting of the time interval, one may take advantage of the parallel architecture of modern computers to speed up the time integration similar to multiple shooting methods \cite{Lions_2001aa}. The pseudo code of the resulting Parareal algorithm is shown in Alg.~\ref{algo:para}. 

\begin{figure}[!t]
	\removelatexerror
	\begin{algorithm}[H]
		\SetKwProg{Parfor}{parfor}{ do}{end}
		\SetKw{KwOr}{or}
		init: $\mathbf{U}_0^{(k)}\leftarrow\mathbf{u}_0$ (for\,all\,$k$) 
		and
		$\bar{\mathbf{u}}^{(0)}_{j}\!, \tilde{\mathbf{u}}^{(0)}_{j}\!\leftarrow\mathbf{0}$ (for\,all\,$j$)\;
		set counter: $k\leftarrow1$\;
		\While{$k\leq2$ \KwOr $\max_j \|{\mathbf{U}}_{j}^{(k)}-{\mathbf{U}}_{j}^{(k-1)}\|>\text{tol}$}{
		\For{$j\leftarrow1$ \KwTo $\numcpu$}{
			solve coarse: $\bar{\mathbf{u}}^{(k)}_{j}\!\leftarrow\mathcal{G}(T_{j},T_{j-1},\mathbf{U}_{j-1}^{(k)})$\;
			post process: $\mathbf{U}^{(k)}_{j}\!\leftarrow\tilde{\mathbf{u}}^{(k-1)}_{j}+\bar{\mathbf{u}}^{(k)}_{j}-\bar{\mathbf{u}}^{(k-1)}_{j}$\;
		}
		\Parfor{$j\leftarrow1$ \KwTo $\numcpu$}{
			solve fine: $\tilde{\mathbf{u}}^{(k)}_{j}\leftarrow\mathcal{F}(T_{j},T_{j-1},\mathbf{U}_{j-1}^{(k)})$\;
		}
		increment counter: $k\leftarrow k+1$\;
		}
		\caption{Parareal as proposed in \cite{Lions_2001aa}\label{algo:para}.}
	\end{algorithm}
\end{figure}

\subsection{Discussion of the Algorithm}
The algorithm solves two kinds of problems in a nested loop until convergence is reached: a cheap problem defined on a coarse time and possibly spatial grid is solved sequentially (line 5, Alg.~\ref{algo:para}) to propagate missing initial conditions and high-fidelity problems are solved in parallel on the intervals $\mathcal{I}_j$ (line 9, Alg.~\ref{algo:para}). For both problems the implicit Euler method \eqref{eq:euler} can chosen, or alternatively a higher order method.

The solution of the cheap problem at $T_{j}$ is denoted by
\begin{align}\label{eq:coarse}
	\bar{\mathbf{u}}_{j}=\mathcal{G}(T_{j},T_{j-1},{\mathbf{U}}_{j-1})
\end{align}
which is computed by propagating the initial value ${\mathbf{U}}_{j-1}$ from $T_{j-1}$ to $T_{j}$ by coarsely discretizing \eqref{eq:daes1} in time, i.e., using large time steps $\Delta t$. The solution of the high-fidelity problem is
\begin{align}\label{eq:fine}
	\tilde{\mathbf{u}}_{j}=\mathcal{F}(T_{j},T_{j-1},{\mathbf{U}}_{j-1})
\end{align}
obtained by solving \eqref{eq:daes1} with initial condition ${\mathbf{U}}_{j-1}$ using fine discretizations, i.e., small $\delta t$. This allows to rewrite the update equation~\eqref{eq:para} in Alg.~\ref{algo:para} (line 6) as
$$ \mathbf{U}^{(k)}_{j}=\tilde{\mathbf{u}}^{(k-1)}_{j}+\bar{\mathbf{u}}^{(k)}_{j}-\bar{\mathbf{u}}^{(k-1)}_{j}.
$$

It can be shown, \cite[Theorem 1]{Gander_2008aa}, that Alg.~\ref{algo:para} yields the correct solution until time $T_k$ after $k$ iterations, so the correct solution is obtained after at most $\numcpu$ iterations. This implies that Parareal does not take longer than the sequential time stepping procedure using the fine solver $\mathcal{F}$ from \eqref{eq:fine} if neglecting the computational costs of the coarse solution operator $\mathcal{G}$. However, in this case Parareal requires up to $\numcpu$-times more CPU time due to its parallel processing.

\subsection{Discussion of the Algebraic Equations}
For the eddy current problem line~6 of Alg.~\ref{algo:para} must be adapted to reflect the differential algebraic character, i.e.
$$ \mathbf{P}\mathbf{U}^{(k)}_{j}=\mathbf{P}\tilde{\mathbf{u}}^{(k-1)}_{j}+\mathbf{P}\bar{\mathbf{u}}^{(k)}_{j}-\mathbf{P}\bar{\mathbf{u}}^{(k-1)}_{j}
$$
with a subsequent solve of \eqref{eq:constraint} to obtain a consistent $\mathbf{Q}\mathbf{u}^{(k)}_{j}$. However, when using Implicit Euler as shown in \eqref{eq:euler}, this step is automatically taken care of. Similarly, the norm in line~3 should be adapted to only account for differential components, e.g. by considering a projection or the eddy current losses.

\section{Numerical Example}\label{sec:num}
\begin{figure}
	\centering
	\includegraphics[width=.44\textwidth]{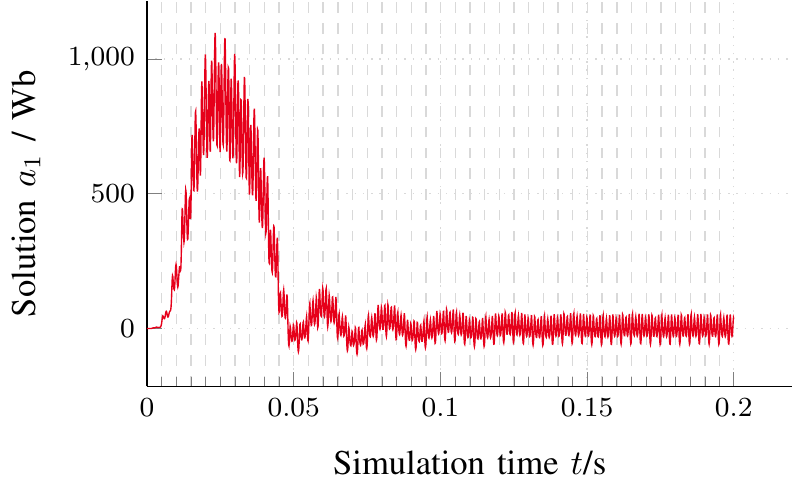}
	\caption{Time stepping dynamics of the discrete magnetic vector potential}
	\label{fig:sol}
\end{figure}

Parareal is particularly interesting for problems with multi-tone solutions, e.g. due to slotting as present in the machine "im\_3kw" as visible in Fig.~\ref{fig:sol}. This was also observed in molecular-dynamics \cite{Baffico_2002aa}. However, speed-ups have also been observed for less favorable problems \cite{Gander_2008aa}.

The algorithm was implemented in GNU Octave \cite{Eaton_2015aa}. GetDP is used for the simulation of the 2D model with 8308 degrees of freedom \cite{Geuzaine_2007aa,Gyselinck_2001aa}. They are executed on an Intel Xeon cluster with $80\times2.00$GHz cores, i.e., $8\times$E7-8850 and $1$TB DDR3 memory.  

\subsection{Sequential Time Steppers}
As sequential time stepper the implicit Euler scheme is applied for 10 electrical periods, i.e., $\mathcal{I}=(0, 0.2]$\,s. The time grid of this simulation is refined from $\delta t=10^{-3}$\,s and $\delta t=10^{-4}$\,s to $\delta t=10^{-5}$\,s. The coarsest sequential simulation takes approx. 15min, while the ones with finer grids correspondingly more time, i.e., 2h and 20h, respectively. Each time step requires ca. $0.3$s for matrix reassembly, Newton and solving the systems.

\subsection{Parareal}
\begin{figure}
	\centering
	\includegraphics[width=.45\textwidth]{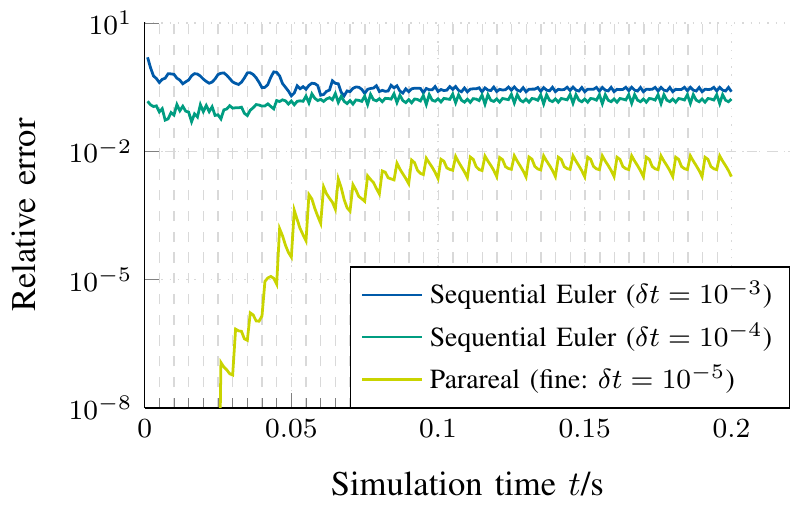}
	\caption{Relative $l_2$ error w.r.t. classical backward Euler ($\delta t=10^{-5}$); dashed vertical lines show the 40 intervals of size $h=0.05$\,s on which the problem is solved in parallel in the case of Parareal (green)}
	\label{fig:errors}
\end{figure}

The Parareal implementation uses OpenMP parallelized calls of GetDP. The implicit Euler method is used with time step sizes $\delta t=10^{-5}$\,s and $\Delta t=10^{-3}$\,s for the fine and coarse problem, respectively. Fig.~\ref{fig:errors} shows the errors in comparison with the sequential reference simulation at $\delta t=10^{-5}$\,s. 
For the Parareal simulation $\numcpu=40$ cores have been used. After $k=4$ iterations, a relative $l_2$ accuracy of $10^{-2}$ has been obtained. The sudden increase of the error after approx. $T_4$ can be explained by the convergence of the Parareal algorithm: it is expected to reproduce the sequential time-stepper's solution for $t<T_k$ ($k=4$).
The computation has a potential speed-up of $\numcpu/k=10$ with respect to the reference simulation when neglecting communication costs and the coarse grid solution. The speed-up is obtained since the effective length of the time interval was reduced by a factor of $\numcpu=40$,  but iterated $k=4$ times. The parallelization allows to obtain errors below $1$\% within the same time that a sequential simulation needs to get errors of $100$\%. However, due to suboptimal implementation the actual speed-up was only ca. 3 times.

\section{Conclusion}\label{sec:conclusion}
This paper proposes the usage of Parareal for eddy current problems and discussed necessary modification due to non-conducting regions in the computational domain. A first implementation shows quick convergence of the iterative algorithm and promises a high speed-up. Future research will investigate the optimal choice of the coarse propagator.  

\section*{Acknowledgement}
This work was supported by DFG grants SCHO 1562/1-1 and CL 143/11-1, the Excellence Initiative of German Federal and State Governments and the Graduate School for CE at TU Darmstadt. The authors would like to thank R.V. Sabariego and N. Marsic for the fruitful discussions on GetDP.

\end{document}